\documentclass[pdflatex,sn-mathphys-num]{sn-jnl}

\usepackage{graphicx}%
\usepackage{multirow}%
\usepackage{amsmath,amssymb,amsfonts}%
\usepackage{amsthm}%
\usepackage{mathrsfs}%
\usepackage[title]{appendix}%
\usepackage{xcolor}%
\usepackage{float}
\usepackage{textcomp}%
\usepackage{manyfoot}%
\usepackage{booktabs}%
\usepackage{algorithm}%
\usepackage{algorithmicx}%
\usepackage{algpseudocode}%
\usepackage{listings}%
\usepackage{tabularx}             
\usepackage{array}
\usepackage{booktabs}
\usepackage{makecell}

\newcolumntype{L}{>{\raggedright\arraybackslash}X}   
\newcolumntype{Y}{>{\centering\arraybackslash}X}     

\newcolumntype{Y}{>{\centering\arraybackslash}X}  
\newcolumntype{L}{>{\raggedright\arraybackslash}X}  

\algblock{ParallelFor}{EndParallelFor}
\algnewcommand\algorithmicparallelfor{\textbf{parallel for}}
\algnewcommand\algorithmicendparallelfor{\textbf{end parallel for}}
\algrenewtext{ParallelFor}[1]{\algorithmicparallelfor\ #1\ \algorithmicdo}
\algrenewtext{EndParallelFor}{\algorithmicendparallelfor}

\theoremstyle{thmstyleone}%
\theoremstyle{thmstyletwo}%

\theoremstyle{thmstylethree}%

\begin{document}

\title[Article Title]{PanDelos-plus: A parallel algorithm for computing sequence homology in
pangenomic analysis}


\author[1]{\fnm{Simone} \sur{Colli}}\email{simone.colli@studenti.unipr.it}
\author[1]{\fnm{Emiliano} \sur{Maresi}}\email{emiliano\_maresi@hotmail.it}
\author[1]{\fnm{Vincenzo} \sur{Bonnici}}\email{vincenzo.bonnici@unipr.it}


\affil[1]{\orgdiv{Department of Mathematical, Physical and Computer Sciences}, \orgname{University of Parma}, \orgaddress{\street{Parco Area delle Scienze 53/a (Campus)}, \city{Parma}, \postcode{43124}, \country{Italy}}}



\abstract{

The identification of homologous gene families across multiple genomes is a
central task in bacterial pangenomics traditionally requiring computationally
demanding all-against-all comparisons.
PanDelos addresses this challenge with an alignment-free and parameter-free
approach based on k-mer profiles, combining high speed, ease of use,
and competitive accuracy with state-of-the-art methods.
However, the increasing availability of genomic data requires tools that can
scale efficiently to larger datasets.
To address this need, we present PanDelos-plus, a fully parallel, gene-centric
redesign of PanDelos.
The algorithm parallelizes the most computationally intensive phases
(Best Hit detection and Bidirectional Best Hit extraction) through
data decomposition and a thread pool strategy, while employing
lightweight data structures to reduce memory usage.
Benchmarks on synthetic datasets show that PanDelos-plus
achieves up to 14x faster execution and reduces memory usage by up to
96\%, while maintaining accuracy.
These improvements enable population-scale comparative genomics to be
performed on standard multicore workstations, making large-scale bacterial pangenome
analysis accessible for routine use in everyday research.
}


\maketitle

\newpage
\section*{Introduction}\label{sec:introduction}

With the rapid progress of high-throughput sequencing projects,
biological data is growing exponentially, creating a need for efficient
and scalable algorithms for further analysis. \cite{yeh2023high}
This need is particularly evident in pangenomics, a discipline that has emerged
to study the entire repertoire of gene families in genomes of a given clade.
In order to understand the genetic diversity within a species, an
abstract structure called pangenome is used. A pangenome is built by identifying groups
of homologous genes. \cite{vernikos2015ten, muthamilarasan2019multi, kim2020current}

The pangenome is a collection of all the genes present in a set of genomes
divided into core, accessory and singleton genes. Core genes, shared by all
genomes, encode essential cellular functions.
Dispensable genes occur in only part of the genomes and often provide adaptive
advantages in specific environments. Singleton genes, unique to a single
genome, usually arise from horizontal gene transfer or reflect
organism-specific adaptations \cite{vernikos2015ten}

Approaches to pangenome content discovery should take into account that gene
mutations, which can occur during gene replication and transitions
may introduce sequence alterations. \cite{brittnacher2011pgat, contreras2013get_homologues, benedict2014itep, chaudhari2016bpga}

Mutations, along with strong evolutionary selection of core genes even if
transmitted with few modifications, make the task of recognizing homologous
genes difficult. In fact, the number of variations that affect dispensable
genes varies and the similarity between homologous sequences tends to decrease
depending on their phylogenetic distance. \cite{bonnici2018pandelos}

The importance of pangenome information is revealed in terms of clinical
applications. Pangenomic studies have been helpful in the identification of
drug targets in vaccines, allowing the targeting of possible vaccine candidates
and antibacterials, helping researchers to decipher their virulence mechanisms
\cite{serruto2009genome, muzzi2007pan}, to detect strain-specific virulence
factors \cite{d2010legionella}, to distinguish between lineage and niche-specific
bacterial population \cite{zhang2014pan} and to investigate pathogens in
epidemic diseases \cite{holt2008high}.

Computationally, retrieval of a pangenome is an NP-hard problem \cite{nguyen2015building},
mainly due to all-against-all comparisons between gene sets.
For this reason, the analysis of pangenomes is a complex and time-consuming task.
So far, many tools have been developed by combining different techniques,
algorithms, and thresholds.

A variety of bioinformatic tools have been developed to address the challenge
of pangenome reconstruction, each with distinct methodologies for gene
clustering, leading to different trade-offs in terms of speed, accuracy, and
scalability. Among the state-of-the-art used tools are Roary \cite{page2015roary},
Panaroo \cite{tonkin2020producing}, PPanGGOLiN \cite{gautreau2020ppanggolin},
and PanDelos \cite{bonnici2018pandelos}.

Roary is widely recognized for its speed, particularly when dealing with
complete high-quality genomes, but its performance can be highly sensitive
to assembly errors present in the input genomes. Panaroo is designed to
specifically address and correct errors often found in draft genome
assemblies, making it suitable for less refined datasets. PPanGGOLiN aims for
scalability across a large number of genomes, though it often requires careful
tuning of its parameters to achieve optimal results. Finally, PanDelos
distinguishes itself by excelling in speed through its alignment-free approach.
It is also able to automatically determine the optimal k-mer value using a
'context-aware parameter inference' method, thus eliminating the need for
manual parameter tuning. This makes it particularly effective in
scenarios that involve high phylogenetic diversity.

Benchmarks have demonstrated that single-core PanDelos has demonstrated
significant advantages in this landscape, outperforming other tools. \cite{bonnici2021challenges}
And thanks to its architecture PanDelos is part of new methodologies like PanDelos-frags,
that infers missing genetic information and thus manages incomplete genomes. \cite{bonnici2023pandelos}

Despite the advancements offered by these tools, many existing solutions are
not fully optimized for analyzing very large datasets.
They often rely on single-threaded algorithms or employ memory-intensive data
structures, or both, making the analysis of large datasets a time-consuming and
resource-intensive task. PanDelos, while being a fast and accurate tool, is not
scalable due to its memory-intensive data structures, which limits its
ability to handle larger datasets efficiently.

We introduce PanDelos-plus, a re-engineered, parallelized version of PanDelos
designed for high-throughput pangenomic analysis.
This work leverages a multi-threading approach and optimized,
low-memory data structures to enable the rapid and efficient analysis
overcoming the scalability limitations of PanDelos.

In this work, we present an evaluation of PanDelos-plus on both real dataset
(including Escherichia coli, Salmonella enterica, Mycoplasma and Xanthomonas
campestris) and synthetic collections generated with PANPROVA (a benchmarking tool that simulates
prokaryotic pangenome evolution starting from a complete ancestral genome) \cite{bonnici2022panprova}.
comprising up to 600 genomes.
Results show up to a 14x speedup and a memory reduction of up to 96\%
compared to PanDelos, while preserving accuracy.

\section*{Basic notions}\label{sec:basic_notion}

Given a dataset consisting of a list of $n$ genomes $\mathbb{G} = \{G_1,\ G_2, \ldots,\ G_n\}$,
each genome $G_i$ is represented by a list of $m$ genes
$G_i = \{g_{i,1},\ g_{i,2}, \ldots,\ g_{i,m}\}$. The number of genes in the
genome $G_i$ is denoted as $|G_i|$.

A gene $g_{i,j} \in \Gamma ^*$ is a string $s = a_1a_2\ldots a_n$ constructed
over a nucleotide or amino acid alphabet $\Gamma$.
For a sequence $s = a_1a_2a_y\ldots a_w\ldots a_n$, with $y < w \le n$, the
substring of $s$ starting at position $y$ and ending at position $w$ is
denoted by $s[y,\ w]$.

Given a fixed value $k$ it's possible to identify a k-mer, defined as a
string $\alpha \in \Gamma^*$ such that $|\alpha| = k$.
Inside a gene $g$ represented by the sequence $s$ of length, defined
as the number of characters inside the sequence, denoted as $|s|$ can
occurs $|s|-k-1$ k-mers.

For a fixed length $k$, the k-mer dictionary of a gene $g$, represents all
k-mers contained inside the gene, and is defined as:
\begin{equation}
D_k(g) = \{\alpha \in \Gamma^* : \exists \rho \text{ such that } g[\rho,\ \rho + k] = \alpha,\ 1 \le \rho \le |g|\}.
\label{eq:kmer_dictionary}
\end{equation}
where $k$ is the length of the k-mers, $\alpha$ is a string built using the
alphabet $\Gamma$, and $g[\rho,\ \rho + k]$ denotes the substring of
$g$ starting at position $\rho$ and ending at position $\rho + k$.

Each k-mer can appear multiple times within a gene $g$. The set of occurrences
of a k-mer $\alpha$ in $g$ is:
\begin{equation}
occ_g(\alpha) = \{\rho : 1 \le \rho \le |g| \land g[\rho,\ \rho + k] = \alpha\}.
\label{eq:kmer_occurrences}
\end{equation}
where $k$ is the length of the k-mers, $\alpha$ is a string built using the
alphabet $\Gamma$, and $g[\rho,\ \rho + k]$ denotes the substring of
$g$ starting at position $\rho$ and ending at position $\rho + k$.

The multiplicity of a k-mer $\alpha$ in a gene $g$, denoted $mult_g(\alpha)$,
is the total number of occurrences of $\alpha$ in $g$:
\begin{equation}
mult_g(\alpha) = |occ_g(\alpha)|.
\label{eq:kmer_multiplicity}
\end{equation}

Dictionary-based approach relies on the k-mers, which are highly sensitive to
the length $k$ of the words that compose the dictionary. The best resolution
of $k$ in pangenomic analysis is theoretically shown to be the best as follow
by the formula:
\begin{equation}
k =\log_{|\Gamma|}{\sum_{i=1}^n |G_i|}
\label{eq:k_length}
\end{equation}

where $G_i$, is the i-th genome of the dataset $\mathbb{G}$,
$|\Gamma|$ is the size of the alphabet that is considered, and $n$ is the
number of genomes in the dataset. According to the studies reported in \cite{bonnici2016informational, manca2017principles}.

For each pair of genes $g_1 \in G_i$ and $g_2 \in G_j$, the sequence similarity is computed
using the generalized Jaccard similarity on k-mers multiplicity, defined as:

\begin{equation}
J_k(g_1, g_2) = \frac{
    \sum_{w \in D_k(g_1) \cap D_k(g_2)} min(mult_{g_1}(w), mult_{g_2}(w))
}{
    \sum_{w \in D_k(g_1) \cup D_k(g_2)} max(mult_{g_1}(w), mult_{g_2}(w))
}
\label{eq:jaccard_k}
\end{equation}
where $D_k(g_1)$ and $D_k(g_2)$ are the k-mer dictionaries (as defined in
Equation \eqref{eq:kmer_dictionary}) of the genes $g_1$ and $g_2$, respectively,
multiplicities $mult_{g_1}(w)$ and $mult_{g_2}(w)$ are the respective counts of
the k-mer $w$ in each gene (as defined in Equation \eqref{eq:kmer_multiplicity})
and $min$ and $max$ are the standard minimum and maximum functions between
the two values.

There is no theory that defines a non-empirical threshold to be applied to the
Jaccard similarity in the context of pangenomic analysis. Therefore, we introduce a
filtering criteria based on the relative length of the gene pairs under consideration.
The filtering criterion is based on a length-based threshold $\theta$, such that $0 \le \theta \le 1$,
that is applied to the gene pairs to be compared.
Given two genes $g_i$ and $g_j$, are considered similar enough to be compared if
\begin{equation}
|g_i| \ge \theta \cdot |g_j|\ and\ |g_j| \ge \theta \cdot |g_i|
\label{eq:length_threshold}
\end{equation}
where $|g_i|$ and $|g_j|$ are the lengths of the genes $g_i$ and $g_j$, respectively.

This length-based threshold \(\theta\) is applied to exclude gene pairs that
are considered too dissimilar in size, under the assumption that highly
unbalanced lengths imply low sequence similarity.
By applying this filter prior to the computation of the Jaccard similarity,
we reduce the number of unnecessary comparisons and improve overall efficiency.

In addition to the length-based threshold, we introduce a second threshold,
denoted as $\tau$, which is applied after the computation of the Jaccard similarity.
This threshold is also not empirically defined, but is theoretically motivated
by considerations on k-mer coverage and overlap.
To explain the rationale: from a sequence $s$, it is possible to extract at
most $|s|/k$ distinct non-overlapping k-mers.
If the average multiplicity of k-mers is close to 1, this quantity
approximates $1/k$  of all k-mer occurrences in $s$.
However, a set of non-overlapping k-mers lacks the positional information
needed to reconstruct the original sequence.
This ensures that the similarity is not only high, but also reciprocal, which
is a desirable property in homolog detection, is introduced an additional
filtering step based on the directional Jaccard similarity defined as:

\begin{equation}
p_k(s \to t) = \frac{\sum_{w \in D_k(s \cap t)} occ_s(w)}{|s| - k + 1}.
\label{eq:directional_jaccard}
\end{equation}
where $occ_s(w)$ is the number of occurrences of the k-mer $w$ in the sequence
$s$ (as defined in Equation \eqref{eq:kmer_occurrences}).

Two genes $g_i$ and $g_j$ are considered homologous candidates only if both
directional Jaccard similarities exceed this threshold
$p_k (g_i \to g_j) \ge \tau \land p_k (g_j \to g_i) \ge \tau$

To identify homologous genes between two genomes, $G_i$ and $G_j$, the Jaccard
similarity is first computed for all possible pairs of their respective genes.
From these similarity scores, the Best Hit (BH) can be determined.
A gene's Best Hit is defined as the set of genes in the opposing genome that
share the highest Jaccard similarity.
Formally, for a gene $g_{i,l}$ in the genome $G_i$, the Best Hit with the respective genome $G_j$ is defined as:
\begin{equation}
BH(g_{i,l}, G_j) = \{g_{j,k} \in G_j : J_k(g_{i,l}, g_{j,k}) = max_{g_{j, k} \in G_j} J_k(g_{i,l}, g_{j,k})\}
\label{eq:best_hit}
\end{equation}
where $g_{i,l}$ is the $l$-th gene of the genome $G_i$, $g_{j,k}$ is the $k$-th gene of the genome $G_j$,
and $J_k(g_{i,l}, g_{j,k})$ is the Jaccard similarity between these two genes.

The Bidirectional Best Hit is a pair of genes that are the Best Hits of each other. So can be defined as:
\begin{equation}
BBH(g_{i, l}, G_j) = \{ g_{j,k} \in G_j : g_{j,k} \in BH(g_{i,l}, G_j) \land g_{i,l} \in BH(g_{j,k}, G_j) \}
\label{eq:bidirectional_best_hit}
\end{equation}
where $g_{i,l}$ is the $l$-th gene of the genome $G_i$, $g_{j,k}$ is the $k$-th gene of the genome $G_j$, and
$BH(g_{i,l}, G_j)$ and $BH(g_{j,k}, G_i)$ are the Best Hits (as defined in Equation \eqref{eq:best_hit}).

\section*{Methodology}\label{sec:methodology}


PanDelos-plus is a parallelized version of PanDelos, designed to enhance the scalability of the pangenomic analysis process.
The proposed algorithm compares genomes by evaluating similarity between individual genes, treating each gene as a basic unit
of comparison. This gene-centric approach enables the inference of homology relationships through pairwise
gene comparisons, which can be efficiently parallelized.
The most computationally demanding steps are the Best Hit detection and Bidirectional Best Hit extraction,
which are designed to leverage parallel computing via data decomposition.
Parallelization is implemented using a thread pool: each comparison task is submitted as
an independent unit of work, and threads iteratively process available tasks from the pool until all computations are completed.

\subsection*{PanDelos methodology and limitations}

PanDelos is a parameter-free, alignment-free algorithm for identifying gene homology
in pangenomic analysis. It leverages on information theory and network analysis
to achieve this. PanDelos quantifies homology using k-mer multiplicity,
with the k-length determined in a non-empirical manner. \cite{bonnici2018pandelos}

The process of detecting gene homology in PanDelos is organized into five key
steps that integrate candidate selection using k-dictionaries with a refinement
phase based on network analysis. Initially, an optimal k-mer
length is determined based on the characteristics of the input genome collection.
Next, genes are compared, and potential homologous pairs are identified.
This selection begins by requiring a minimum overlap between the
k-dictionaries of two genes. Following this, the generalized Jaccard similarity
is calculated to evaluate the similarity between genes, enabling the
identification of bidirectional Best Hits. These hits form a homology network,
from which gene families are extracted after a refinement process.
\cite{bonnici2018pandelos}

The core algorithm of PanDelos' pipeline relies on the enhanced suffix
array \cite{abouelhoda2002enhanced} to compute sequence similarity.
This structure, combines a suffix array \cite{manber1993suffix} with the longest common prefix
(LCP) array, drastically improve the efficiency of substring comparisons and overall
runtime performance. The suffix array lists all the suffixes of a string in lexicographic order,
each associated with its starting position, while the LCP array captures the length of the longest
common prefix between adjacent suffixes. This combination allows PanDelos to
perform fast and accurate similarity detection. \cite{bonnici2018pandelos}

Despite its significant impact on speed, this approach has a notable drawback,
because the suffix array is memory-intensive, making it the most
resource-demanding component of the pipeline. This limitation is caused by the
need to store a large amount of data in memory, particularly the LCP array,
which can grow significantly for large genomes.

\subsection*{PanDelos-plus pipeline overview}

PanDelos-plus pipeline is designed as revisitation of PanDelos, so some steps of
the pipeline are inherited from PanDelos.
The pipeline is composed of three macro phases. The first phase (a) is used to
determine the optimal k-mer length basing on the input dataset.
The second phase (b) focuses on an all-against-all comparison of all input
genomes, to generate Bidirectional Best Hit.
The third phase (c) is used to extract gene families from the bidirectional
Best Hit.

The most demanding phase is the second one (b). This phase is made up of three
sub-phases: (i) \textit{Multiplicity Vector building},
(ii) \textit{Best Hit detection}, and (iii)
\textit{Bidirectional Best Hit extraction}.
The first sub-phase (i) is a serial and covers all analyzed genomes,
and the second and third sub-phases (ii and iii) are parallel and consists in an
all-against-all comparison of genomes.

The multiplicity vector building phase (i) is serial. During this phase
all possible k-mers are extracted and associated to an unique integer (i.1).
Then k-mers with same unique integer are grouped together and sorted in ascending order
in order to produce a multiplicity vector (i.2).
The multiplicity vector is a data structure associated with each gene
that consists of an ordered vector of pairs. Each pair consists of a k-mer,
represented as a unique identifier, and the respective multiplicity inside
the associated gene.
The vector is sorted in ascending order according to the first element of
the pair.

The Best Hit detection phase (ii) is parallelized via data decomposition.
During this phase the algorithm computes an all-against-all comparison
of the genes of the pair of genome considered.
This phase compute the similarity, between all possible pair of genes, using
the generalized Jaccard similarity, and compute Best Hits on rows. Computed
Best Hits are stored inside a jagged array data structure.

The Bidirectional Best Hit extraction phase (iii) is parallelized. During this
phase similarity values and jagged array are used to extract candidate columns
to become the bidirectional Best Hits.

\begin{figure}[H]
    \centering
    \includegraphics[width=\textwidth,height=0.9\textheight,keepaspectratio]{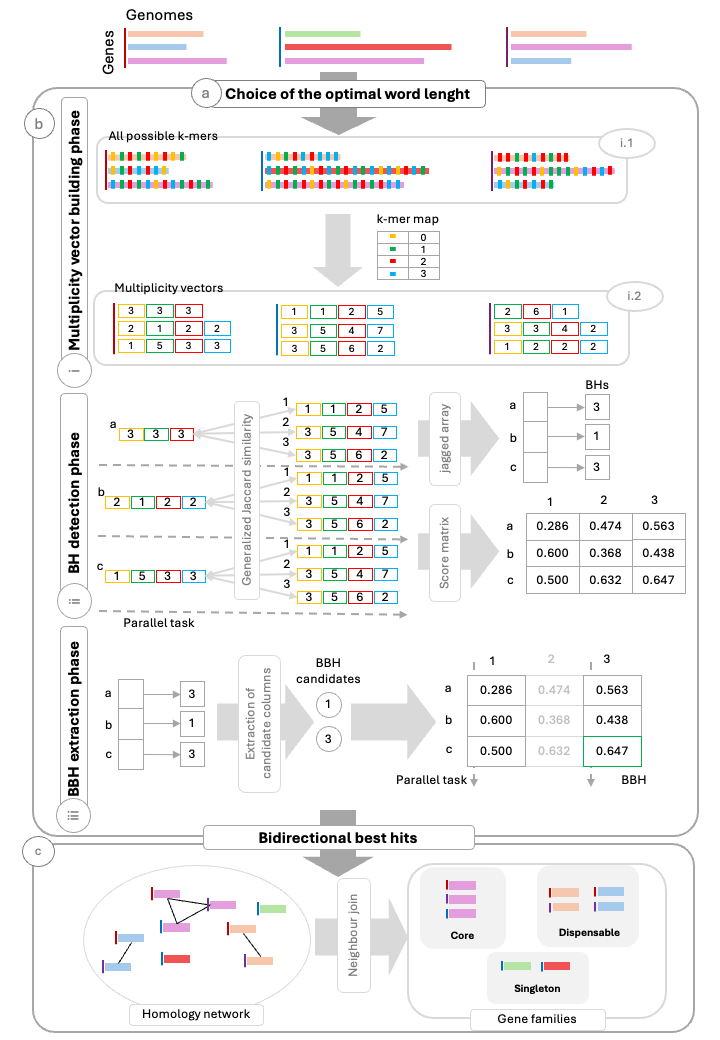}
    \caption{
        PanDelos-plus pipeline (revisiting PanDelos) in three stages: (a)
        automatic selection of the optimal k-mer length; (b) all-against-all
        gene comparison to generate Bidirectional Best Hits (BBH), with three
        subphases—serial construction of multiplicity vectors
        (k-mer extraction and grouping), parallel Best Hit detection via
        generalized Jaccard similarity (stored in a jagged array), and
        parallel BBH extraction; (c) final clustering into gene families
        based on the BBH.
    }
    \label{fig:pdp-overview}
\end{figure}

\subsection*{Multiplicity vector building phase}

The first phase of the algorithm is the construction of the multiplicity vectors (MV).
The MV is a data structure that all genes will handle in order to store all different k-mers,
along with related multiplicities.

Consider a function $f(\alpha) : \alpha \to n$, such that $\alpha \in \Gamma^*$ and $n \in \mathbb{N}$,
that is used to associate each kmer to a unique positive integer.

Given $\alpha_1$, $\alpha_2$ and $\alpha_3$, such that
$\alpha_1,\ \alpha_2, \alpha_3 \in \Gamma^*$ with $\alpha_1 \ne \alpha_2$ 
and $\alpha_1 = \alpha_3$, the function ensures that
$f(\alpha_1) = f(\alpha_3)$ whenever $f(\alpha_1) \neq f(\alpha_2)$

The MV contains a set of pairs $(f(\alpha),\ mult_{g_{i,j}}(\alpha))$,
where $f(\alpha)$ is the unique identifier of the k-mer $\alpha$ and $mult_{g_{i,j}}(\alpha)$
is the multiplicity of the k-mer $\alpha$ in the gene $g_{i,j}$.

The MV is defined as:
\[
MV(g_{i,j}) = (f(\alpha),\ mult_{g_{i,j}}(\alpha)) : \alpha \in D_k(g_{i,j})\}
\]

The MV is sorted in way that the first element is the one with the lower $f(\alpha_1)$
value and the last one is the one with the higher $f(\alpha_2)$, so that
$f(\alpha_1) < f(\alpha_2) \land \alpha_1 \neq \alpha_2$.

\begin{algorithm}[H]
\caption{Multiplicity Vector building phase for a single gene}
\label{alg:mv-building}
\begin{algorithmic}[1]
\Require Gene $g_{i,j}$, k-mer length $k$, alphabet $\Gamma$
\Ensure Multiplicity Vector $MV(g_{i,j})$
\State Initialize an empty ordered map $M$ \Comment{key: k-mer id, value: multiplicity}
\For{$p \gets 1$ to $|g_{i,j}| - k + 1$} \Comment{Slide a window of length $k$}
    \State $\alpha \gets g_{i,j}[p \ldots p+k-1]$ \Comment{Extract substring of length $k$}
    \State $id \gets f(\alpha)$ \Comment{Compute unique identifier for $\alpha$}
    \If{$id \in M$}
        \State $M[id] \gets M[id] + 1$ \Comment{Increment multiplicity}
    \Else
        \State Insert $(id, 1)$ into $M$ \Comment{First occurrence of this k-mer}
    \EndIf
\EndFor
\State $MV(g_{i,j}) \gets$ convert $M$ to a vector of pairs $(id, multiplicity)$ in ascending order of $id$
\State \Return $MV(g_{i,j})$
\end{algorithmic}
\end{algorithm}

The work defined in Algorithm \ref{alg:mv-building} is done for each
gene in the genomes $G_i$ and $G_j$.
Any type of computation that could include any type of string comparison
has been substituted by a computation on the MV data structure.

\newpage
\subsection*{Best Hit detection phase}

The second phase of the algorithm is the Best Hit detection.
The Best Hit detection is a parallelized process that computes the similarity between
the genes of two genomes $G_i$ and $G_j$ and detect Best Hits.

All possible comparisons can be represented as a matrix, where the rows represent the genes of genome $G_i$
and the columns represent the genes of genome $G_j$.
The matrix is defined as:
\[
S_{i,j} = \begin{bmatrix}
    s_{1,1} & s_{1,2} & \ldots & s_{1,m} \\
    s_{2,1} & s_{2,2} & \ldots & s_{2,m} \\
    \vdots & \vdots & \ddots & \vdots \\
    s_{n,1} & s_{n,2} & \ldots & s_{n,m}
\end{bmatrix}
\]

Where $m = |G_j|$, $n = |G_i|$, and $s_{l,k}$ is the similarity value
between the gene $g_{i,l}$ and the gene $g_{j,k}$, with $1 \le l \le n \land 1 \le k \le m$.

In addition to this matrix, a jagged array is defined to keeps tracks of the
Best Hit on each row.
The jagged array is defined as:

\[
A_{i,j} = \begin{bmatrix}
    Ar_1 \\
    Ar_2 \\
    \vdots \\
    Ar_n
\end{bmatrix}
\]

Where $Ar_i$ is a vector that contains the columns identifiers of Best Hits
of the $i$-th row of the matrix $S_{i,j}$.
So the $Ar_i$ is defined as:
\[
Ar_i =
    \begin{bmatrix}
        a_{i,1} & a_{i,2} & \ldots & a_{i,k} \\
    \end{bmatrix}
\]
where $j_{i,k}$ represents the index of the column of the $k$-th Best Hit of the $i$-th row.

Using this type of representation each row of the matrix $S_{i,j}$ and the corresponding
row of the jagged array $A_{i,j}$ can be considered independent from each other.

It is possible to parallelize the computation process, by splitting the
matrix $S_{i,j}$ into $n$ independent unit of work (UoW), where each unit of
work is represented by a row of the matrix. Each UoW includes computing the
similarity between the gene $g_{i,l}$, $1 \le l \le m$ and all the genes of
the genome $G_j$, along with tracking the Best Hits on the rows. The number of
units of work is equal to the number of genes in the first genome $G_i$.
The procedure is detailed in Algorithm \ref{alg:best-hit-detection}.

\begin{algorithm}[H]
\caption{Best Hit detection phase}
\label{alg:best-hit-detection}
\begin{algorithmic}[1]
\Require Genomes $G_i$, $G_j$
\Ensure Similarity matrix $S_{i,j}$ and jagged array $A_{i,j}$
\State Initialize $S_{i,j}$ as an $n \times m$ matrix \Comment{$n = |G_i|$, $m = |G_j|$}
\State Initialize $A_{i,j}$ as an empty jagged array of size $n$
\ParallelFor{$g_{i,l} \in G_i$} \Comment{Each row is an independent unit of work}
    \ForAll{$g_{j,k} \in G_j$}
        \State $s_{l,k} \gets similarity(g_{i,l}, g_{j,k})$
        \State Store $s_{l,k}$ in $S_{i,j}[l,k]$
    \EndFor
    \State $Ar_l \gets$ indices of maximum values in row $S_{i,j}[l,*]$ \Comment{Best Hits of $g_{i,l}$}
    \State Store $Ar_l$ in $A_{i,j}[l]$
\EndParallelFor
\State \Return $S_{i,j}, A_{i,j}$
\end{algorithmic}
\end{algorithm}

\subsection*{Bidirectional Best Hit extraction phase}

The third phase of the algorithm is the Bidirectional Best Hit extraction.
The Bidirectional Best Hit extraction is a parallelized process
that extract the bidirectional Best Hits starting
from the $S_{i,j}$ and the $A_{i,j}$ matrices.

To identify a Bidirectional Best Hit is required that the Best Hit of the $i$-th row
of the $S_{i,j}$ matrix is the same as the Best Hit of the $j$-th column.
So just only a part of the matrix $S_{i,j}$ is needed to be considered in that phase.
The part of the matrix that is needed to be considered is the one that contains
at least one Best Hit of the $i$-th row of the $S_{i,j}$ matrix on the $j$-column.

Taking advantage of the jagged array $A_{i,j}$, it is possible to identify the
relevant columns (columns that contains at least one Best Hit) that need to
be considered in the computation of the Bidirectional Best hits.
Like the Best Hit detection phase the Bidirectional Best Hit extraction phase
is parallelized leveraging on the matrix representation and data independency,
by splitting the matrix $S_{i,j}$ into $m$ independent columns.
The number of independent columns is equal to the number of genes
in the second genome $G_j$ less the number of column that never appear in the $A_{i,j}$.

Relying on the independency of the columns and the matrix representation of
the computed Jaccard similarities during the Best Hit detection phase, each
column is considered as single, and independent, UoW.
Each UoW includes the computation of the Best Hit on the $k$-th column
of the $S_{i,j}$ matrix and the cross check for the bidirectional Best Hits
with the $A_{i,j}$ matrix Best Hits.
The Bidirectional Best Hit extraction procedure is summarized in Algorithm
\ref{alg:bbh-extraction}.

\begin{algorithm}[H]
\caption{Bidirectional Best Hit extraction phase}
\label{alg:bbh-extraction}
\begin{algorithmic}[1]
\Require Similarity matrix $S_{i,j}$, jagged array $A_{i,j}$, genomes $G_i$, $G_j$
\Ensure Set of Bidirectional Best Hits $BBH$
\State $BBH \gets \emptyset$
\State Identify relevant columns $k$ of $S_{i,j}$ from $A_{i,j}$ \Comment{Columns that contain at least one Best Hit}
\ParallelFor{each relevant column $k$} \Comment{Each column is an independent UoW}
    \ForAll{rows $l$ such that $k \in A_{i,j}[l]$} \Comment{Candidate Best Hits}
        \If{$S_{i,j}[l,k] = \max_{r} S_{i,j}[r,k]$} \Comment{$g_{i,l}$ is a Best Hit of $g_{j,k}$}
            \If{$l \in A_{j,i}[k]$} \Comment{Cross-check: reciprocal Best Hit}
                \State Add pair $(g_{i,l}, g_{j,k})$ to $BBH$
            \EndIf
        \EndIf
    \EndFor
\EndParallelFor
\State \Return $BBH$
\end{algorithmic}
\end{algorithm}

\subsection*{Dataset construction}\label{sec:datasets}

Two sets of experiments have been conducted to evaluate the performance
of PanDelos-plus. The first set, refered as real dataset, is composed of
real bacterial genomes, and is used to compare the performance of PanDelos-plus
with the original PanDelos algorithm, and to measure scalability in runtime
and memory usage as the thread count grows.
The second set, referred as synthetic dataset, is composed of synthetic
bacterial genomes, and is used to measure scalability on large genome
collections generated via PANPROVA.

PANPROVA is a computational tool designed to simulate
the evolution of prokaryotic pangenomes by evolving the complete genomic
sequence of an ancestral isolate to generate synthetic datasets.
It operates by taking a full genome and its gene annotations,
then simulating evolutionary events (such mutations, gene loss, gene duplication,
and horizontal gene transfer) under user-defined parameters, resulting in a
phylogenomic tree populated with fully assembled, evolutionarily related
synthetic genomes.
This tool is used primarily to create realistic synthetic bacterial genomic
datasets for benchmarking and evaluating pangenomic analysis tools, especially
in scenarios where complete and controlled reference data are needed.

The real dataset \ref{tab:rd-overview} includes four real-world bacterial groups taken from
the PanDelos original paper.
These groups (supplementary material) comprise complete genomes of Escherichia coli, Salmonella enterica, and
Xanthomonas campestris species, along with a collection of different
Mycoplasma complete genomes in REFSEQ annotation.
\begin{table}[htbp]
    \centering
    \caption{Summary of the real bacterial datasets used for benchmarking PanDelos-plus.
    Each dataset consists of complete genomes downloaded in REFSEQ format from NCBI,
    originally used in the PanDelos study.
    The table reports, for each species group, the total number of genomes, the total gene count,
    and descriptive statistics of gene content per genome (maximum, minimum, mean, and standard deviation).
    }
    \label{tab:rd-overview}
    \begin{tabularx}{\textwidth}{L Y Y Y Y Y Y}
        \toprule
        Dataset   & Number of genomes & Total number of genes & Max. number of genes
                  & Min. number of genes & Mean of number of genes & Standard deviation of number of genes \\
        \midrule
        Echerichia coli & 10 & 46724 & 5195 & 4209 & 4672.40 & 340.41 \\
        Salmonella enterica & 7 & 29898 & 4325 & 4212 & 4271.14 & 35.71 \\
        Xanthomonas campestris & 14 & 55290 & 4300 & 3023 & 3949.29 & 339.58 \\
        Mycoplasma & 64 & 48117 & 1495 & 479 & 751.83 & 181.41 \\
        \bottomrule
    \end{tabularx}
\end{table}

The synthetic dataset \ref{tab:sd-overview} includes a collection of synthetic bacterial pangenome datasets
generated using PANPROVA starting from an ancestral genome.
Used genomes are generated starting from an ancestral genome of Escherichia coli and evolving it using PANPROVA
with an htg pool of 9 genomes (supplementary material).

These datasets comprises 12 collections, from $50$ to $600$, created by increasing
the number of genomes by $50$ at each step maintaining the previous genomes untouched.
Keeping the previous genomes untouched, it is possible to evaluate the
scalability of PanDelos-plus as both the number of genomes and the total
gene count grow, without introducing artifacts from new genome generation.

\begin{table}[htbp]
    \centering
    \caption{Overview of synthetic datasets used in the experiments.
    Each dataset is generated using PANPROVA, starting from an ancestral genome
    of Escherichia coli by increasing the number of genomes by $50$ at each step
    without introducing artifacts from new genome generation.}
    \label{tab:sd-overview}
    \begin{tabularx}{\textwidth}{L Y Y Y Y Y Y}
        \toprule
        Dataset   & Number of genomes & Total number of genes & Max. number of genes 
                  & Min. number of genes & Mean of number of genes & Standard deviation of number of genes \\
        \midrule
        Dataset 1 & 50 & 269882 & 5559 & 5202 & 5397.64 & 80.27 \\
        Dataset 2 & 100 & 540231 & 5608 & 5202 & 5402.31 & 75.79 \\
        Dataset 3 & 150 & 812378 & 5650 & 5202 & 5415.85 & 83.70 \\
        Dataset 4 & 200 & 1084178 & 5651 & 5202 & 5420.89 & 86.53 \\
        Dataset 5 & 250 & 1356037 & 5658 & 5202 & 5424.15 & 87.42 \\
        Dataset 6 & 300 & 1628181 & 5687 & 5202 & 5427.27 & 89.85 \\
        Dataset 7 & 350 & 1901281 & 5707 & 5202 & 5432.23 & 92.27 \\
        Dataset 8 & 400 & 2173975 & 5707 & 5202 & 5434.94 & 91.45 \\
        Dataset 9 & 450 & 2446401 & 5758 & 5202 & 5436.45 & 92.77 \\
        Dataset 10 & 500 & 2719951 & 5758 & 5202 & 5439.90 & 94.68 \\
        Dataset 11 & 550 & 2994448 & 5762 & 5202 & 5444.45 & 96.73 \\
        Dataset 12 & 600 & 3268551 & 5790 & 5202 & 5447.59 & 98.37 \\
        \bottomrule
    \end{tabularx}
\end{table}

\newpage
\section*{Results}\label{sec:results}

All the tests have been performed on the high-performance computing (HPC) cluster
of the University of Parma.
Tests involved node with 2 AMD EPYC 7282 2.8 GHz 16c CPU, 32 cores, and 256
GB of RAM.

The tests cover both real and synthetic datasets (subsection \ref{sec:datasets}),
measuring wall-clock time, peak memory usage, and speedup efficiency as the thread count varied.

\subsection*{Real dataset results}

In all four datasets, PanDelos-plus, thanks to his lightweight data structure
and parallelization of the most demanding phases, significantly outperforms the
original PanDelos in both execution time and memory consumption (Table
\ref{tab:pdp-pd-all-comparison}).

\begin{table}[htbp]
    \centering
    \caption{Comparison of PanDelos and PanDelos-plus performance metrics across real bacterial
    datasets. Each dataset analysed using PanDelos with a single thread and PanDelos-plus
    with 32 threads. Metrics include execution time, peak memory usage,
    speedup factor, and percentage reduction in memory consumption.}
    \label{tab:pdp-pd-all-comparison}
    \begin{tabularx}{\textwidth}{L Y Y Y Y Y Y}
        \toprule
        Dataset & \multicolumn{2}{c}{\makecell{PanDelos\\(1 thread)}} & \multicolumn{2}{c}{\makecell{PanDelos-plus\\(32 threads)}} & \makecell{Runtime\\speedup} & \makecell{Memory\\reduction\\(\%)} \\
        \cmidrule(lr){2-3}\cmidrule(lr){4-5}\cmidrule(lr){6-6}\cmidrule(lr){7-7}
        & Time (s) & Memory (GB) & Time (s)  & Memory (GB) & Speedup & Reduction \\
        \midrule
        Escherichia coli & 619 & 20.2 & 46 & 0.7 & 13.5 & 96.5 \\
        Salmonella enterica & 216 & 12.8 & 16 & 0.5 & 13.5 & 96.1 \\
        Xanthomonas campestris & 1015 & 15.7 & 73 & 0.7 & 13.9 & 95.5 \\
        Mycoplasma & 563 & 2.2 & 65 & 0.7 & 8.7 & 68.2 \\
        \bottomrule
    \end{tabularx}
\end{table}

PanDelos-plus demonstrates significant performance advantages, in execution
time and memory consumption over PanDelos across all datasets.

For instance, on the Escherichia coli dataset, PanDelos required $619 s$ and $20.2 GB$ of RAM,
whereas PanDelos-plus completed the same analysis in just $46 s$ using only
$0.7 GB$, achieving a 13.5x speedup and a $96.5 \%$ reduction in memory.

In the Mycoplasma test, execution time fell from $563 s$ to $65 s$ and peak RAM
from $2.2 GB$ to $0.7 GB$, corresponding to an 8.7x acceleration and
$68.2 \%$ less memory usage.

For Salmonella enterica, PanDelos-plus reduced runtime from $216 s$ to $16 s$
and memory from $12.8 GB$ to $0.5 GB$ again a 13.5x speedup alongside a
$96.1 \%$ RAM saving.

Finally, on Xanthomonas campestris, the enhanced implementation ran in $73 s$
instead of $1015 s$ and used $0.7 GB$ rather than $15.7 GB$, translating into a
13.9x faster execution with a $95.5 \%$ reduction in memory footprint.

These substantial gains in both speed and memory efficiency establish
PanDelos-plus as a significantly more powerful tool than its predecessor.
To further understand its performance profile, particularly how effectively
it utilizes computational resources when parallelized, we conducted a strong scaling
analysis.
This analysis, detailed below, examines the impact of an increasing number
of processing threads on execution time and memory usage for a constant
problem size across the same datasets.

To further investigate how effectively PanDelos-plus exploits computational resources
when parallelized, a strong scaling analysis has been conducted.

For all data sets, an almost ideal speed up is observed, as doubling the number
of threads approximately halves the execution time.
Examining the impact of increasing the number of threads on execution time and
memory usage for a fixed problem size across all datasets.

The analysis shows that execution time consistently decreases as the number of
threads increases, exhibiting near-ideal scalability, where doubling the number
of threads nearly halves the runtime.

\begin{figure}[H]
    \centering
    \includegraphics[width=0.8\textwidth,height=0.8\textheight]{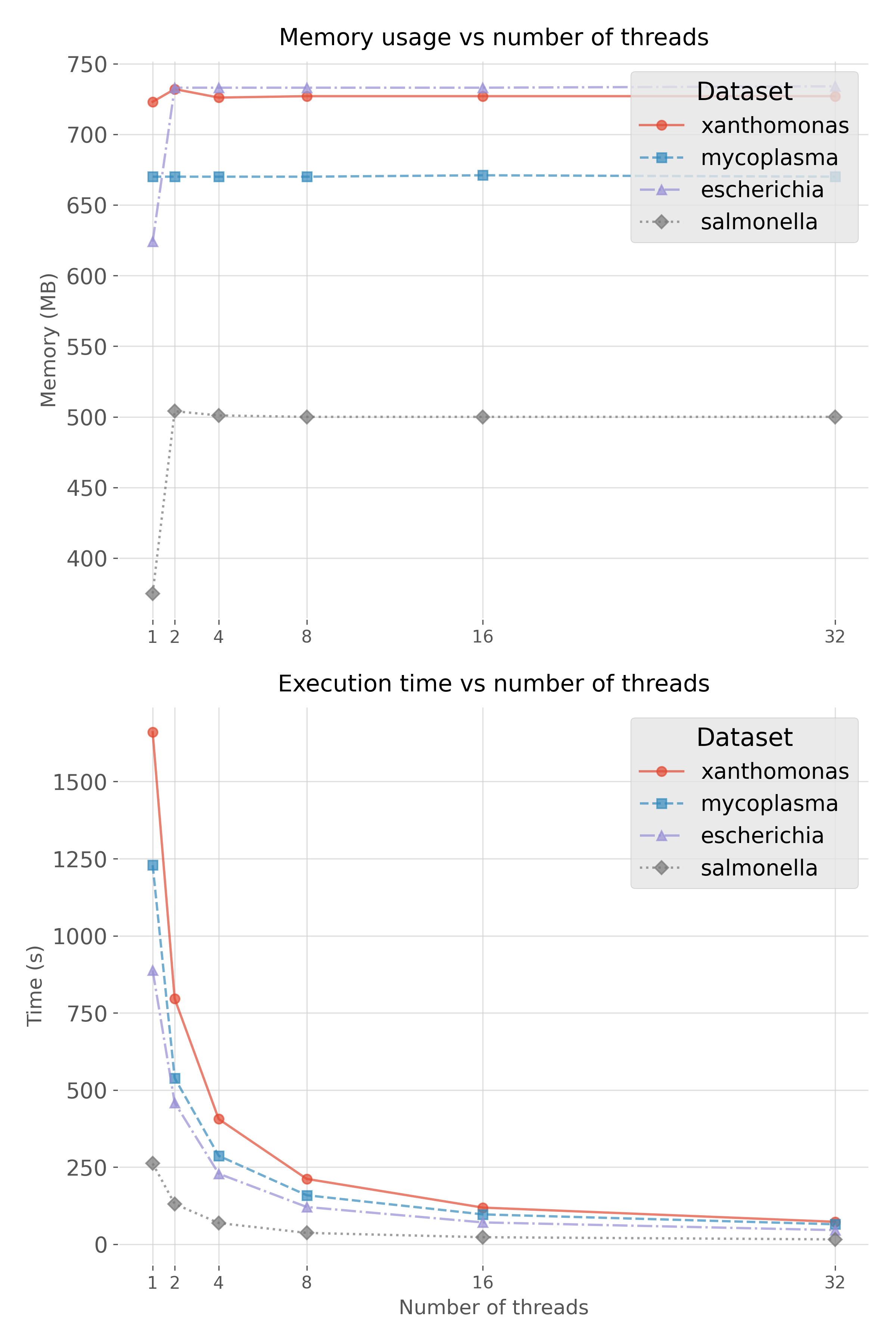}
    \caption{Strong scaling of PanDelos-plus on real datasets.
            Top: Peak memory usage (MB) remains nearly constant as the number of threads
            increases, indicating minimal overhead from parallelization.
            Bottom: Execution time (s) decreases almost inversely with the number of threads,
            confirming efficient parallel scalability across all datasets.
    }
    \label{fig:pdp-threads-plot}
\end{figure}

The plot (Figure \ref{fig:pdp-threads-plot}, top), titled “Memory usage vs number of threads,” shows that memory
consumption (in MB), barring a few minor variations, remains essentially constant
for each dataset as the number of threads increases.
This indicates that increasing parallelism does not significantly increase the
memory footprint for a given problem size. The plot
(Figure \ref{fig:pdp-threads-plot}, bottom), titled “Execution time vs number
of threads,” shows a significant reduction in execution time
(in seconds) as the number of threads increases for all datasets.

\subsection*{Synthetic dataset results}

To assess the scalability of PanDelos-plus on increasingly large datasets, have
been conducted two series of tests on the synthetic dataset generated using PANPROVA.

\begin{figure}[H]
    \centering
    \includegraphics[width=0.8\textwidth,height=0.8\textheight]{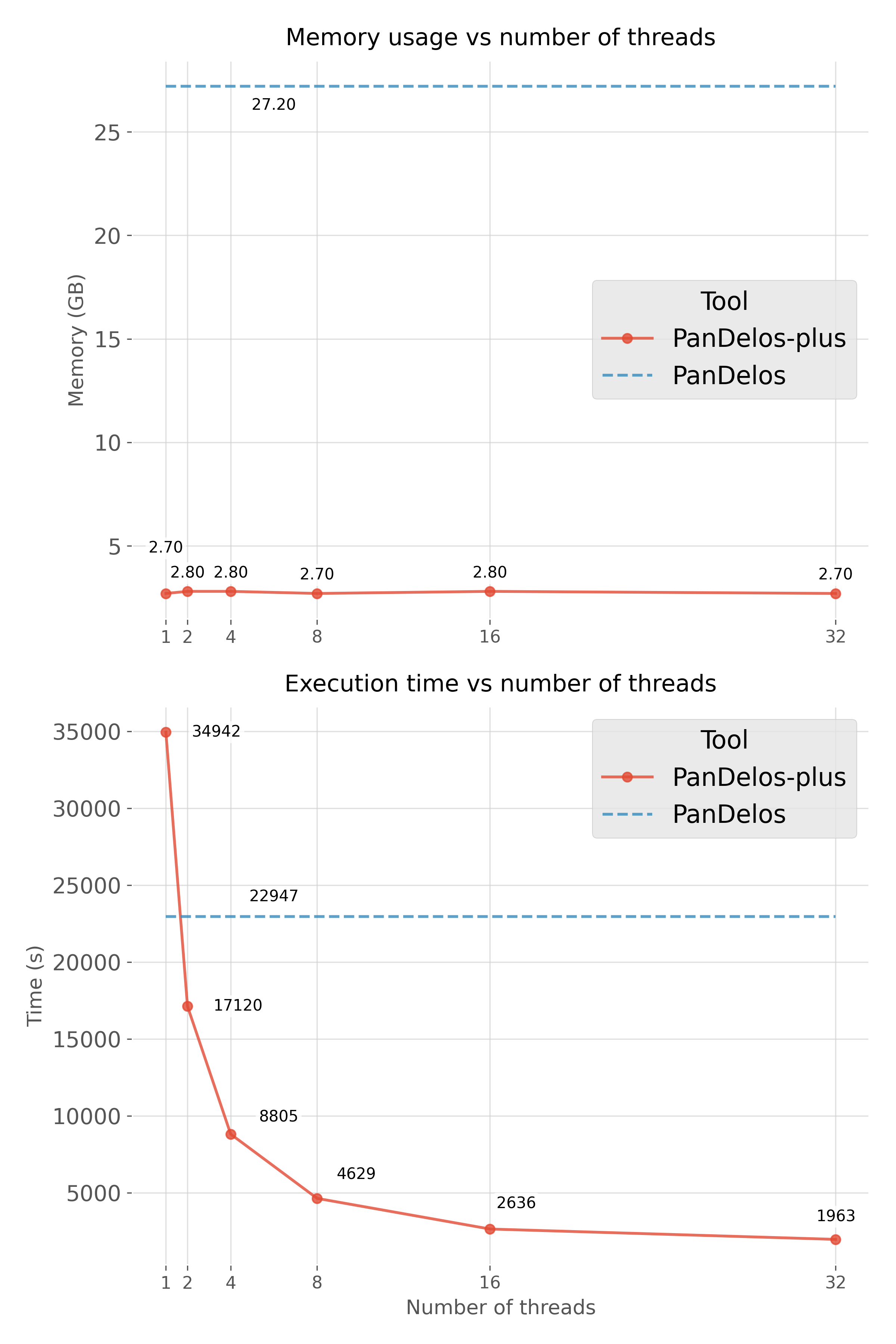}
    \caption{Scaling behavior of PanDelos-plus on a synthetic dataset of 50 genomes.
            Top: Memory usage (GB) remains almost constant as the number of threads
            increases, while PanDelos shows much higher consumption $27.2 GB$ vs $2.8 GB$.
            Bottom: Execution time (s) decreases almost linearly with thread count,
            from approximately $34942 s$ (9.6 h) with 1 thread to about $1963 s$ (0.5 h) with 32 threads,
            achieving up to 18x acceleration compared to the single-threaded execution.}
    \label{fig:pdp-sts-plot}
\end{figure}

\begin{figure}[H]
    \centering
    \includegraphics[width=0.8\textwidth,height=0.8\textheight]{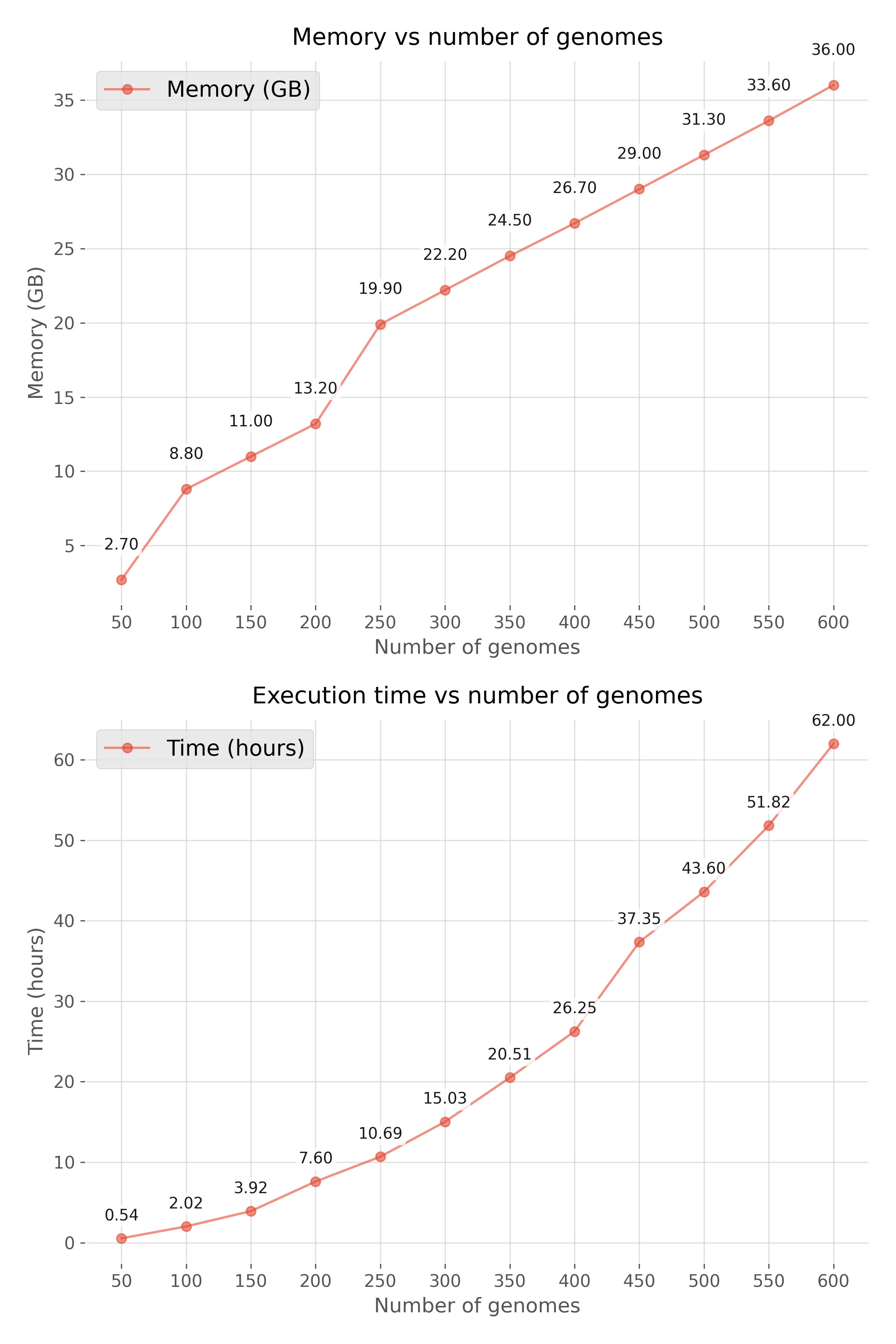}
    \caption{Scalability of PanDelos-plus with increasing dataset size ($50-600$ genomes) at $32$ threads.
            Top: Peak memory usage grows approximately linearly with the number of genomes,
            remaining below $40 GB$ even for $600$ genomes.
            Bottom: Execution time scales sublinearly with dataset size, increasing
            from $\approx 0.5$ h ($50$ genomes) to $\approx 62$ h ($600$ genomes),
            indicating near-linear time complexity and effective parallel efficiency even
            on very large datasets.
    }
    \label{fig:pdp-sts-plot-number-of-genomes}
\end{figure}

In the first one (Figure \ref{fig:pdp-sts-plot}), the dataset size has been fixed
to $50$ genomes (Table \ref{tab:sd-overview}, Dataset $1$)
and varied the number of threads from $1$ to $32$, measuring peak memory usage
and execution time. In this instance PanDelos has been introduced as a baseline
for comparison, using a single thread as it does not support parallel execution.
In the second part (Figure \ref{fig:pdp-sts-plot-number-of-genomes}), has been
used a constant number of threads (for instance, $32$) and observed how
memory usage and execution time grow as the number of genomes
increases from $50$ to $600$ (Table \ref{tab:sd-overview}, Datasets $1-12$).
Here, PanDelos has not been included as it cannot handle datasets of this size.

When analysing the $50$ genomes (Table \ref{tab:sd-overview},
Dataset $1$), peak memory usage remains essentially constant and the time
decreases significantly as the number of threads increases.
This indicates that increasing parallelism does not significantly affect the
RAM required for a fixed problem size, while it does improve execution time.

The plot (Figure \ref{fig:pdp-sts-plot}, top), titled “Memory usage vs number of threads”,
shows that memory consumption (in GB) remains essentially constant
as the number of threads increases, with only minor variations.
Here PanDelos-plus's memory usage is significantly lower than PanDelos's, respectively
approximately $2.8 GB$ vs approximately $27.2 GB$

The plot (Figure \ref{fig:pdp-sts-plot}, bottom), titled “Execution time vs number of threads”,
shows a significant reduction in execution time (in seconds) as the number of threads
increases. PanDelos-plus takes about $34942 s$ ($\approx 9.6 h$) with a
single thread to analyze the $50$ genomes, resulting slower than PanDelos's
$22947 s$ ($\approx 6.37 h$) for the same dataset. But
as threads are added, execution time decreases almost linearly:
with $2$ threads it drops to roughly $17120 s$ ($\approx 4.7 h$), with $4$
threads to $8805 s$ ($\approx 2.4 h$), with $8$ threads to $4629 s$
($\approx 1.3 h$), with $16$ threads to $2636 s$ ($\approx 0.7 h$),
and finally with $32$ threads to $\approx 1963 s$ ($\approx 0.5 h$).
Since the execution with $2$ threads PanDelos-plus is already faster than
PanDelos. The behavior observed confirms that, for a moderately sized dataset,
PanDelos-plus effectively leverages 32 cores, reducing the execution time by
approximately 18x compared to single-threaded execution.

The next test, using a constant number of threads ($32$)
shows that both memory usage and execution time grow as the number of genomes
increases from $50$ to $600$.

The plot (Figure \ref{fig:pdp-sts-plot-number-of-genomes}, top),
titled “Memory usage vs number of genomes”, shows that with $32$ threads
PanDelos-plus’s peak memory consumption (in GB) grows as
the number of genomes increases from 50 to 600.
For $50$ genomes, the peak memory is approximately $2.7 GB$.
As we add more genomes, memory usage rises almost linearly reaching $\approx 36 GB$
at $600$ genomes.
This nearly linear trend indicates that PanDelos-plus allocates memory
in direct proportion to the total gene count, remaining within acceptable
bounds (under $40 GB$) even for a $600$-genome pangenome on a $32$-core node.

The plot (Figure \ref{fig:pdp-sts-plot-number-of-genomes}, bottom),
titled “Execution time vs number of genomes”, shows that
with $32$ threads the runtime increases from approximately $0.5 h$ ($1900 s$)
for $50$ genomes up to approximately $62 h$ ($225000 s$) for $600$ genomes.
The plot exhibits a sublinear relationship: although the number of
genomes grows by a factor of $12$ (from 50 to 600), execution time also
increases by roughly 12x, going from $\approx 0.5 h$ to $ \approx 62 h$.
This trend indicates that PanDelos-plus maintains near-linear time complexity
with respect to the number of genomes when run with $32$ threads,
leveraging internal optimizations and parallelism to keep
analysis times reasonable even on very large datasets.

\section*{Conclusions}\label{sec:conclusion}

PanDelos-plus represents a substantial enhancement over the original PanDelos
algorithm, overcoming key limitations in scalability, runtime efficiency, and
memory consumption. Leveraging a gene-centric approach, the use of lightweight
data structures, and thread-level parallelism, PanDelos-plus achieves significant
improvements in runtime and RAM usage over its predecessor.

Benchmarks on real and synthetic datasets confirm that the redesigned
implementation drastically reduces execution time and memory requirements, achieving
up to 14x faster performance and up to $96\%$ less memory usage.
Importantly, the analysis of synthetic datasets shows that while the
single-threaded version of PanDelos remains faster than PanDelos-plus
when only one core is used, the parallel design of PanDelos-plus allows
it to surpass PanDelos already with two threads, achieving nearly ideal
scalability as the number of threads increases.

These findings demonstrate that PanDelos-plus efficiently exploits modern
multicore architectures, maintaining stable memory usage while reducing runtime
almost proportionally to the number of threads. Furthermore, its near-linear growth
in both execution time and memory with increasing dataset size highlights its
robust scalability, enabling the analysis of hundreds of genomes within practical
timeframes.

Overall, PanDelos-plus extends the applicability of the PanDelos framework to
population-scale comparative genomics, making large-scale pangenome analysis
feasible on standard multicore workstations and HPC environments.

\section*{Acknowlegment}\label{sec:acknowlegment}

This research benefited from the High-Performance Computing facility of the
University of Parma, Italy (HPC.unipr.it).

We gratefully acknowledge the support of the CINI InfoLife laboratory in this research.

This project has been partially funded by the University of Parma (Italy), project number FIL\_2024\_PROGETTI\_B\_BONNICI.


\end{document}